\documentclass[published]{JHEP} 

\usepackage{epsfig}

\def\beq{\begin{equation}}
\def\eeq{\end{equation}}
\def\beqa{\begin{eqnarray}}
\def\eeqa{\end{eqnarray}}
\def\lla{\left\langle}
\def\rra{\right\rangle}
\def\za{\alpha}
\def\zb{\beta}
\def\lsim{\mathrel{\raise.3ex\hbox{$<$\kern-.75em\lower1ex\hbox{$\sim$}}} }
\def\gsim{\mathrel{\raise.3ex\hbox{$>$\kern-.75em\lower1ex\hbox{$\sim$}}} }
\newcommand{\Rbs}{\mbox{${{\scriptstyle \not}{\scriptscriptstyle R}}$}}

\title{$LR$ Scalar Mixings and One-loop Neutrino Masses } 
\author{Otto C. W. Kong\\
Institute of Physics, Academia Sinica, Nankang, Taipei TAIWAN 11529} 

\abstract{
Within the framework of the complete theory of supersymmetry without 
R-parity, where all possible R-parity violating terms are admitted,
we perform a systematic analytical study of all sources of
neutrino masses up to ``direct one-loop" (defined explicitly below) 
level. In the passing, we present the
full result for squark and slepton masses. In particular, there are
interesting $LR$ squark and slepton mixings, which involve both
bilinear and trilinear R-parity violating  parameters. The existence
and important phenomenological implications of such terms have
been largely overlooked in previous studies. In particular, in the studies
under which either one type of the couplings is assumed to vanish
or neglected, the terms would not show up. The $LR$ mixings play a 
central role in neutrino mass generation. Our results look straight
forward to be obtained, which, in our opinion, is an illustration of the
effectiveness of our formulation adopted.
} 


\preprint{rev. Aug 2000}

\begin{document}

\section{Introduction }
The minimal supersymmetric standard model (MSSM) is no doubt the most
popular candidate theory for physics beyond the Standard Model (SM). The
alternative theory with a discrete symmetry called R-parity not imposed
deserves no less attention. In particular, the latter  admits
neutrino masses, without the need for any extra matter field beyond 
the minimal spectrum. At the present time, experimental results from 
neutrino physics\cite{nu} is actually the only data we have demanding 
physics beyond the SM, while signals from supersymmetry (SUSY) are still 
absent. The neutrino data provides strong hints for the existence of 
Majorana type masses. The latter means lepton number violation, which is
suggestive of R-parity violation. Hence, it is easy to appreciate the
interest in R-parity violating (RPV) contributions to neutrino masses.
The study of this topic has a long history, starting from Ref.\cite{HS}. 
Two of the notable papers on different aspects of the topic are given in
Ref.\cite{H} and Ref.\cite{NP}, to which readers are also referred
for references to earlier works. More recent works in the subject 
area\cite{Rnu,HDPRV,ok,AL,as1,GH} mainly focus on the fitting of 
the neutrino oscillation data under different scenarios while a 
comprehensive analysis of all the RPV contributions is still missing. 
This paper aims at providing such a picture. 

Like most of the other recent
studies, we will focus on the sub-eV neutrino mass scale suggested
by the Super-Kamiokande atmospheric neutrino data\cite{sK}, though 
most of our results are actually valid for a much larger range of neutrino 
masses. As illustrated below, there is a tree-level but seesaw 
suppressed contribution and some direct 1-loop contributions. Our level of
treatment in this paper stops there, {\it i.e.} we will, in general,
not go into contributions that are expected to be further suppressed. 
There are direct 1-loop contributions which involve a further seesaw
type suppression hidden inside the loop. These are 1-loop diagrams
that would suggest a null result if electroweak (EW) states are used for the 
particles running inside the loop and only a minimal number of mass insertions
is admitted. When one thinks about the exact result to be obtained from 
using mass eigenstates instead, a nonzero result could emerge. It is 
diffcult to give analytical expressions for the mass eigenstate results.
An approximation to the latter could be obtained by  considering
the EW state diagrams with extra mass insertions. In the case that these
mass insertions are RPV, it typically means extra seesaw type suppression.
We refer to contributions from such diagrams as pseudo-direct
1-loop contributions, which we will discuss without giving explicit
formulae. We list also the well-known results. The idea here is to 
perform a systematic analysis and present the exhaustive list of all
contributions up to the level of treatment. 

A similar comprehensive listing of neutrino mass contributions up to the 
1-loop level (direct or indirect) has been presented in Ref.\cite{kias}. 
However, the latter analysis is limited to a scenario where the ``third
generation couplings dominate". This amounts to admitting only non-zero
$\lambda^{\!\prime}_{i\scriptscriptstyle 33}$'s and
$\lambda_{i\scriptscriptstyle 33}$'s among the trilinear RPV couplings, 
though all nonzero bilinear RPV are indeed included by the authors. In our 
opinion, the maximal mixing result from Super-Kamiokande\cite{sK} brings the 
wisdom of ``third generation domination" under question.  
Refs.\cite{ok} and \cite{as1}, for example illustrate how no (family)
hierarchy, or even an anti-hierarchy, among the RPV couplings may be 
preferred. The present analysis handles the complete theory of supersymmetry 
(SUSY) without R-parity, where all kind of RPV terms are admitted without 
bias. We present complete tree-level mass matrices 
for the scalars in this generic scenario. There is
another  major difference between the two studies.  
Ref.\cite{kias} is interested in performing some numerical calculation. 
While the latter is important for explicit fitting of experimental
numbers, much of the physical origin of the neutrino mass contributions 
are hidden under elements of mixing matrices parametrizing the effective
couplings of the neutrinos to squark or slepton mass eigenstates.
We are interested here in illustrating the explicit origin of each
contribution. Hence, we stay with electroweak (EW) state notation and give
diagrammatic as well as analytical expressions of each individual
contribution. Of particular interest here is a new type of 
contribution involving a RPV LR scalar (squark or slepton) mixings, 
which has been larger overlooked by previous authors. We hope that results 
here will be useful for a better understanding the role of each RPV parameter 
and identifying interesting regions of the extensive parameter space.

To study all the RPV contributions in a single consistent framework, 
one needs an effective formulation of the complete theory of SUSY without
R-parity. The latter theory is generally better motivated than {\it ad hoc} 
versions of RPV theories. The large number of new parameters involved,
however, makes the theory difficult to analyze. It has been 
illustrated\cite{k} that an optimal parametrization, called the single-VEV 
parametrization, can be of great help in making the task manageable. 
The effectiveness of the SVP has been explored to perform an extensive
study on the resultant leptonic phenomenology\cite{k}, to identify new type 
of neutrino mass contributions\cite{as1}, and to study a new 
contribution to neutron electric dipole moment at 1-loop level\cite{as4,as6}, 
as well as new sources of contribution to flavor changing neutral 
current processes such as $b\to s\, \gamma\,$\cite{kk} and 
$\mu \to e \,\gamma\,$\cite{as7}. Studies of neutrino masses and mixings 
under the formulation also include Refs.\cite{ok,AL}. 
In fact, neutrino masses contribution is a central aspect
of RPV effects and is likely to provide the most stringent bounds on the
couplings, though many of the bounds obtained depend on assumptions on
interpretation of neutrino data and could be relaxed or removed by simple
extensions of the theory allowing extra sterile neutrino(s).

One-loop neutrino mass generation in SUSY without R-parity typically
involves $LR$ mixings of squarks or slepton. We want to emphasize again
that  squark and slepton mass matrices presented here are complete,
with all source of R-parity violation included. Such results are
explicitly presented for the first time. 
We consider the results to be interesting in their own right. 

In the appendix, we give also an explicit illustration 
that all the VEV's under the SVP may be taken as real, despite the 
existence of complex parameters in the scalar potential; and give some 
important consistence relationships among some of the
parameters involved. These results have not been published before, and 
serve as important background for clarifying some issues on the 
scalar masses and neutrino mass contributions discussed.

\section{Formulation and Notation } 
We summarize our formulation and notation below.
The most general renormalizable superpotential for the supersymmetric
SM (without R-parity) can be written  as
\begin{equation}
W \!\! = \!\varepsilon_{ab}\left[ \mu_{\alpha}  \hat{H}_u^a \hat{L}_{\alpha}^b 
+ h_{ik}^u \hat{Q}_i^a   \hat{H}_{u}^b \hat{U}_k^{\scriptscriptstyle C}
+ \lambda_{\alpha jk}^{\!\prime}  \hat{L}_{\alpha}^a \hat{Q}_j^b
\hat{D}_k^{\scriptscriptstyle C} + 
\frac{1}{2}\, \lambda_{\alpha \beta k}  \hat{L}_{\alpha}^a  
 \hat{L}_{\beta}^b \hat{E}_k^{\scriptscriptstyle C} \right] + 
\frac{1}{2}\, \lambda_{ijk}^{\!\prime\prime}  
\hat{U}_i^{\scriptscriptstyle C} \hat{D}_j^{\scriptscriptstyle C}  
\hat{D}_k^{\scriptscriptstyle C}   ,
\end{equation}
where  $(a,b)$ are $SU(2)$ indices, $(i,j,k)$ are the usual family (flavor) 
indices, and $(\za, \zb)$ are extended flavor index going from $0$ to $3$.
In the limit where $\lambda_{ijk}, \lambda^{\!\prime}_{ijk},  
\lambda^{\!\prime\prime}_{ijk}$ and $\mu_{i}$  all vanish, 
one recovers the expression for the R-parity preserving case, 
with $\hat{L}_{0}$ identified as $\hat{H}_d$. Without R-parity imposed,
the latter is not {\it a priori} distinguishable from the $\hat{L}_{i}$'s.
Note that $\lambda$ is antisymmetric in the first two indices, as
required by  the $SU(2)$  product rules, as shown explicitly here with 
$\varepsilon_{\scriptscriptstyle 12} =-\varepsilon_{\scriptscriptstyle 21}=1$.
Similarly, $\lambda^{\!\prime\prime}$ is antisymmetric in the last two indices
from $SU(3)_{\scriptscriptstyle C}$.

R-parity is exactly an {\it ad hoc} symmetry put in to make $\hat{L}_{0}$,
stand out from the other $\hat{L}_i$'s as the candidate for  $\hat{H}_d$.
It is defined in terms of baryon number, lepton number, and spin as, 
explicitly, ${\mathcal R} = (-1)^{3B+L+2S}$. The consequence is that 
the accidental symmetries of baryon number and lepton number in the SM 
are preserved, at the expense of making particles and superparticles having 
a categorically different quantum number, R-parity. The latter is actually 
not the most effective discrete symmetry to control superparticle 
mediated proton decay\cite{pd}, but is most restrictive in terms
of what is admitted in the Lagrangian, or the superpotential alone.

A naive look at the scenario suggests that the large number of new 
couplings makes the task formidable. However, it becomes quite
manageable with an optimal choice of flavor bases, the SVP\cite{k}. 
In fact, doing phenomenological studies without specifying a choice 
of flavor bases is ambiguous. It is like doing SM quark physics with 18
complex Yukawa couplings instead of the 10 real physical parameters.
In SUSY without R-parity, the choice of an optimal
parametrization mainly concerns the 4 $\hat{L}_\alpha$ flavors.
Under the SVP, flavor bases are chosen such that : 
1/ among the $\hat{L}_\alpha$'s, only  $\hat{L}_0$, bears a VEV
{\it i.e.} $\langle \hat{L}_i \rangle \equiv 0$;
2/  $h^{e}_{jk} (\equiv \lambda_{0jk} =-\lambda_{j0k}) 
=\frac{\sqrt{2}}{v_{\scriptscriptstyle 0}}{\rm diag}
\{m_{\scriptscriptstyle 1},
m_{\scriptscriptstyle 2},m_{\scriptscriptstyle 3}\}$;
3/ $h^{d}_{jk} (\equiv \lambda^{\!\prime}_{0jk}) 
= \frac{\sqrt{2}}{v_{\scriptscriptstyle 0}}{\rm diag}\{m_d,m_s,m_b\}$; 
4/ $h^{u}_{ik}=\frac{v_{\scriptscriptstyle u}}{\sqrt{2}}
V_{\scriptscriptstyle \!C\!K\!M}^{\dag} {\rm diag}\{m_u,m_c,m_t\}$, where 
$v_{\scriptscriptstyle 0}\equiv \sqrt{2} \, \langle \hat{L}_0 \rangle$
and $v_{\scriptscriptstyle u}\equiv \sqrt{2} \,
\langle \hat{H}_{u} \rangle$.
The big advantage here is that the (tree-level) mass 
matrices for all the fermions  do not involve any of the trilinear 
RPV couplings, even though the approach makes  no assumption 
on any RPV coupling, including those from soft SUSY breaking. Moreover,
and all the parameters used are uniquely defined, with the exception of
some removable phases. In fact, the (color-singlet) charged fermion mass 
matrix reduces to the simple form :
\beq \label{mc}
{\mathcal{M}_{\scriptscriptstyle C}} =
 \left(
{\begin{array}{ccccc}
{M_{\scriptscriptstyle 2}} &  
\frac{g_{\scriptscriptstyle 2}{v}_{\scriptscriptstyle 0}}{\sqrt 2}  
& 0 & 0 & 0 \\
 \frac{g_{\scriptscriptstyle 2}{v}_{\scriptscriptstyle u}}{\sqrt 2} & 
 {{ \mu}_{\scriptscriptstyle 0}} & {{ \mu}_{\scriptscriptstyle 1}} &
{{ \mu}_{\scriptscriptstyle 2}}  & {{ \mu}_{\scriptscriptstyle 3}} \\
0 &  0 & {{m}_{\scriptscriptstyle 1}} & 0 & 0 \\
0 & 0 & 0 & {{m}_{\scriptscriptstyle 2}} & 0 \\
0 & 0 & 0 & 0 & {{m}_{\scriptscriptstyle 3}}
\end{array}}
\right)  \; .
\eeq
Each  $\mu_i$ parameter here characterizes directly the RPV effect
on the corresponding charged lepton  ($\ell_i = e$, $\mu$, and $\tau$).
For any set of other parameter inputs, the ${m}_i$'s can 
then be determined, through a simple numerical procedure, to guarantee that 
the correct mass eigenvalues of  $m_e$, $m_\mu$, and $m_\tau$  are obtained 
--- an issue first addressed and solved in Ref.\cite{k}. The latter issue
is especially important when $\mu_i$'s not substantially smaller than
${ \mu}_{\scriptscriptstyle 0}$ are considered. Such an odd scenario is
not definitely ruled out\cite{k}. However, we would concentrate here on
the more popular scenario with only sub-eV neutrino masses and hence
small $\mu_i$'s. Here deviations of the ${m}_i$'s from the mass eigenvalues
are negligible.

\section{Gauginos, Higgsinos, and Neutrinos }
The tree-level mixings among the  gauginos, higgsinos, and neutrinos 
gives rise to a $7\times 7$ neutral fermion mass matrix $\cal{M_N}$:
\begin{equation}
\label{77}
\cal{M_N} = 
\left (\begin{array}{cccc|ccc}
M_{\scriptscriptstyle 1} & 0 & g_{\scriptscriptstyle 1}
 v_{\scriptscriptstyle u}/2 
& -g_{\scriptscriptstyle 1} v_{\scriptscriptstyle 0}/2 & 0 & 0 & 0  \\
0   & M_{\scriptscriptstyle 2} 
& -g_{\scriptscriptstyle 2} v_{\scriptscriptstyle u}/2 
& g_{\scriptscriptstyle 2} v_{\scriptscriptstyle 0}/2 & 0 & 0 & 0  \\
g_{\scriptscriptstyle 1} v_{\scriptscriptstyle u}/2 & 
-g_{\scriptscriptstyle 2} v_{\scriptscriptstyle u}/2 & 0   
& -\mu_{\scriptscriptstyle 0} 
 & -\mu_{\scriptscriptstyle 1} & -\mu_{\scriptscriptstyle 2} 
& -\mu_{\scriptscriptstyle 3} \\
-g_{\scriptscriptstyle 1} v_{\scriptscriptstyle 0}/2 & 
g_{\scriptscriptstyle 2} v_{\scriptscriptstyle 0}/2 & 
-\mu_{\scriptscriptstyle 0} 
&   0 & 0 & 0 & 0\\
\hline
0 & 0 & -\mu_{\scriptscriptstyle 1}   & 0 & 
(m_\nu^o)_{\!\scriptscriptstyle 1\!1} 
& (m_\nu^o)_{\!\scriptscriptstyle 1\!2} & 
(m_\nu^o)_{\!\scriptscriptstyle 1\!3}
\\
0 & 0 &  -\mu_{\!\scriptscriptstyle 2} & 0  & 
(m_\nu^o)_{\!\scriptscriptstyle 
2\!1} 
& (m_\nu^o)_{\!\scriptscriptstyle 22} & 
(m_\nu^o)_{\!\scriptscriptstyle 23} \\
0 & 0 & -\mu_{\!\scriptscriptstyle 3} & 0 & 
(m_\nu^o)_{\!\scriptscriptstyle 31}
& (m_\nu^o)_{\!\scriptscriptstyle 32} & 
(m_\nu^o)_{\!\scriptscriptstyle 33} \\
 \end{array}
\right )\; ,
\end{equation}
whose basis is $(-i\tilde{B}, -i\tilde{W}, 
\tilde{h}_{\scriptscriptstyle u}^0, \tilde{h}_{\scriptscriptstyle d}^0, 
\nu_{\!\scriptscriptstyle L_1},\nu_{\!\scriptscriptstyle L_2 },
\nu_{\!\scriptscriptstyle L_3})$, with $\tilde{h}_{\scriptscriptstyle d}^0$
being the neutral fermion from $\hat{L}_0$. The latter is guaranteed 
to be predominately a neutralino rather than neutrino, as the mass
matrix clearly illustrates. As pointed out above, for small 
$\mu_i$'s the charged fermion states in the $\hat{L}_i$'s are essentially
the physical states of $e$, $\mu$ and $\tau$. Hence, 
$(\nu_{\!\scriptscriptstyle L_1},\nu_{\!\scriptscriptstyle L_2 },
\nu_{\!\scriptscriptstyle L_3})$ are essentially $\nu_e,\nu_\mu,\nu_\tau$.
All entires in the lower-right $3\times 3$ block $(m_\nu^o)$ are, 
of course, zero at tree level.  They are induced via 1-loop contributions 
to be discussed below. Such contributions are the focus of the 
present study. They are referred to here
as direct 1-loop contributions. 

We can write the general mass matrix in the form of block submatrices:
\begin{equation} \label{mnu}
{\cal M_N} = \left( \begin{array}{cc}
              {\cal M}_n & \xi^{\!\scriptscriptstyle  T} \\
              \xi & m_\nu^o \end{array}  \right ) \;,
\end{equation}
where ${\cal{M}}_n$ is the upper-left $4\times 4$ neutralino mass matrix, 
$\xi$ is the $3\times 4$ block, and $m_\nu^o$ is the lower-right 
$3\times 3$ neutrino block in the $7\times 7$ matrix. The resulting (effective) neutrino mass matrix after block diagonalization 
is given by
\begin{equation} \label{ss}
(m_\nu) = - \xi {\cal M}_n^{\mbox{-}1} \xi^{\!\scriptscriptstyle  T} + (m_\nu^o)
 \;.
\end{equation}
Contributions to the first term here starts at tree level, which are, 
however, seesaw suppressed. The second term is the direct contribution,
which, however, enters in only  at 1-loop level. 
We are interested in the small $\mu_i$ scenario, where the tree-level
contribution is not necessarily expected to be  stronger than such direct
1-loop effects. The 1-loop contributions to the $\xi$ and ${\cal M}_n$ 
blocks are likely to have only a secondary effect on $(m_\nu)$.
The latter, to be called indirect 1-loop contributions, are not included 
in the present analysis.

\section{LR-mixings for Squarks and Sleptons  }
The soft SUSY breaking part of the Lagrangian can be written as 
\beqa
V_{\rm soft} &=& \epsilon_{\!\scriptscriptstyle ab} 
  B_{\za} \,  H_{u}^a \tilde{L}_\za^b +
\epsilon_{\!\scriptscriptstyle ab} \left[ \,
A^{\!\scriptscriptstyle U}_{ij} \, 
\tilde{Q}^a_i H_{u}^b \tilde{U}^{\scriptscriptstyle C}_j 
+ A^{\!\scriptscriptstyle D}_{ij} 
H_{d}^a \tilde{Q}^b_i \tilde{D}^{\scriptscriptstyle C}_j  
+ A^{\!\scriptscriptstyle E}_{ij} 
H_{d}^a \tilde{L}^b_i \tilde{E}^{\scriptscriptstyle C}_j   \,
\right] + {\rm h.c.}\nonumber \\
&+&
\epsilon_{\!\scriptscriptstyle ab} 
\left[ \,  A^{\!\scriptscriptstyle \lambda^\prime}_{ijk} 
\tilde{L}_i^a \tilde{Q}^b_j \tilde{D}^{\scriptscriptstyle C}_k  
+ \frac{1}{2}\, A^{\!\scriptscriptstyle \lambda}_{ijk} 
\tilde{L}_i^a \tilde{L}^b_j \tilde{E}^{\scriptscriptstyle C}_k  
\right] 
+ \frac{1}{2}\, A^{\!\scriptscriptstyle \lambda^{\prime\prime}}_{ijk} 
\tilde{U}^{\scriptscriptstyle C}_i  \tilde{D}^{\scriptscriptstyle C}_j  
\tilde{D}^{\scriptscriptstyle C}_k  + {\rm h.c.}
\nonumber \\
&+&
 \tilde{Q}^\dagger \tilde{m}_{\!\scriptscriptstyle {Q}}^2 \,\tilde{Q} 
+\tilde{U}^{\dagger} 
\tilde{m}_{\!\scriptscriptstyle {U}}^2 \, \tilde{U} 
+\tilde{D}^{\dagger} \tilde{m}_{\!\scriptscriptstyle {D}}^2 
\, \tilde{D} 
+ \tilde{L}^\dagger \tilde{m}_{\!\scriptscriptstyle {L}}^2  \tilde{L}  
  +\tilde{E}^{\dagger} \tilde{m}_{\!\scriptscriptstyle {E}}^2 
\, \tilde{E}
+ \tilde{m}_{\!\scriptscriptstyle H_{\!\scriptscriptstyle u}}^2 \,
|H_{u}|^2 
\nonumber \\
&& + \frac{M_{\!\scriptscriptstyle 1}}{2} \tilde{B}\tilde{B}
   + \frac{M_{\!\scriptscriptstyle 2}}{2} \tilde{W}\tilde{W}
   + \frac{M_{\!\scriptscriptstyle 3}}{2} \tilde{g}\tilde{g}
+ {\rm h.c.}\; ,
\label{soft}
\eeqa
where we have separated the R-parity conserving $A$-terms from the 
RPV ones (recall $\hat{H}_{d} \equiv \hat{L}_0$). Note that 
$\tilde{L}^\dagger \tilde{m}_{\!\scriptscriptstyle \tilde{L}}^2  \tilde{L}$,
unlike the other soft mass terms, is given by a 
$4\times 4$ matrix. Explicitly, 
$\tilde{m}_{\!\scriptscriptstyle {L}_{00}}^2$ corresponds to 
$\tilde{m}_{\!\scriptscriptstyle H_{\!\scriptscriptstyle d}}^2$ 
of the MSSM case while 
$\tilde{m}_{\!\scriptscriptstyle {L}_{0k}}^2$'s give RPV mass mixings.

The SVP also simplifies much the otherwise complicated expressions
for the mass-squared matrix of the scalar sectors. Firstly, we will look 
at the squark sectors. The masses of up-squarks obviously have no RPV 
contribution. The down-squark sector, however, has an interesting result. 
We have the mass-squared matrix as follows : 
\beq \label{MD}
{\cal M}_{\!\scriptscriptstyle {D}}^2 =
\left( \begin{array}{cc}
{\cal M}_{\!\scriptscriptstyle LL}^2 & {\cal M}_{\!\scriptscriptstyle RL}^{2\dag} \\
{\cal M}_{\!\scriptscriptstyle RL}^{2} & {\cal M}_{\!\scriptscriptstyle RR}^2
 \end{array} \right) \; ,
\eeq
where
\beqa
{\cal M}_{\!\scriptscriptstyle LL}^2 &=&
\tilde{m}_{\!\scriptscriptstyle {Q}}^2 +
m_{\!\scriptscriptstyle D}^\dag m_{\!\scriptscriptstyle D}
+ M_{\!\scriptscriptstyle Z}^2\, \cos\!2 \beta 
\left[ -\frac{1}{2} + \frac{1}{3} \sin\!^2 \theta_{\scriptscriptstyle W}\right] \; ,
\nonumber \\
{\cal M}_{\!\scriptscriptstyle RR}^2 &=&
\tilde{m}_{\!\scriptscriptstyle {D}}^2 +
m_{\!\scriptscriptstyle D} m_{\!\scriptscriptstyle D}^\dag
+ M_{\!\scriptscriptstyle Z}^2\, \cos\!2 \beta 
\left[  - \frac{1}{3} \sin\!^2 \theta_{\scriptscriptstyle W}\right] \; ,
\eeqa 
and
\beq \label{RL}
({\cal M}_{\!\scriptscriptstyle RL}^{2})^{\scriptscriptstyle T} = 
A^{\!{\scriptscriptstyle D}} \frac{v_{\scriptscriptstyle 0}}{\sqrt{2}}
- m_{\!\scriptscriptstyle D} \, \mu_{\scriptscriptstyle 0}^* \, \tan\!\beta 
- (\, \mu_i^*\lambda^{\!\prime}_{ijk}\, ) \; 
\frac{v_{\scriptscriptstyle u}}{\sqrt{2}} \; .
\eeq
Here, $m_{\!\scriptscriptstyle D}$ is the down-quark mass matrix, 
which is diagonal under the parametrization adopted; 
$(\, \mu_i^*\lambda^{\!\prime}_{ijk}\, )$ denotes the 
$3\times 3$ matrix $(\;)_{jk}$ with elements listed; and 
$\tan\!\beta =\frac{v_{\scriptscriptstyle u}}{v_{\scriptscriptstyle 0}}$.
Note that all the VEV's can be taken as real, so long as the tree level
scalar potential is considered (see the appendix). Apart from the first 
$A^{\!{\scriptscriptstyle D}}$ term, the remaining terms in 
$({\cal M}_{\!\scriptscriptstyle RL}^2)^{\scriptscriptstyle T}$ 
are $F$-term contributions; in particular, the last term gives 
``SUSY conserving" but R-parity violating contributions;
note that the existence of nonzero
$F$-terms or electroweak symmetry breaking VEV's can be interpreted 
as a consequence of SUSY breaking though.  The full $F$-term 
part in the above equation can actually be written together as 
$(\, \mu_{\scriptscriptstyle \za}^*\lambda^{\!\prime}_{{\scriptscriptstyle \za}jk}) 
\, \frac{v_{\scriptscriptstyle u}}{\sqrt{2}}$ where the $\alpha=0$ 
term, which  vanishes for $j \ne k$, gives the second term in the RHS. 
The latter, of course, is just 
the usual $\mu$-term contribution in the MSSM case. 

Next we move on to the slepton sector. From Eq.(\ref{soft}) above, we 
can see that the ``charged Higgs" should be
considered  together with the sleptons. We have hence an
$8\times 8$ mass-squared matrix of the following $1+4+3$ form :
\beq \label{ME}
{\cal M}_{\!\scriptscriptstyle {E}}^2 =
\left( \begin{array}{ccc}
\widetilde{\cal M}_{\!\scriptscriptstyle H\!u}^2 &
\widetilde{\cal M}_{\!\scriptscriptstyle LH}^{2\dag}  & 
\widetilde{\cal M}_{\!\scriptscriptstyle RH}^{2\dag}
\\
\widetilde{\cal M}_{\!\scriptscriptstyle LH}^2 & 
\widetilde{\cal M}_{\!\scriptscriptstyle LL}^{2} & 
\widetilde{\cal M}_{\!\scriptscriptstyle RL}^{2\dag} 
\\
\widetilde{\cal M}_{\!\scriptscriptstyle RH}^2 &
\widetilde{\cal M}_{\!\scriptscriptstyle RL}^{2} & 
\widetilde{\cal M}_{\!\scriptscriptstyle RR}^2  
\end{array} \right) \; ;
\eeq
where
\beqa
\widetilde{\cal M}_{\!\scriptscriptstyle LL}^2 &=&
\tilde{m}_{\!\scriptscriptstyle {L}}^2 +
m_{\!\scriptscriptstyle L}^\dag m_{\!\scriptscriptstyle L}
+ (\mu_{\!\scriptscriptstyle \za}^* \mu_{\scriptscriptstyle \zb})
+ M_{\!\scriptscriptstyle Z}^2\, \cos\!2 \beta 
\left[ -\frac{1}{2} +  \sin\!^2 \theta_{\!\scriptscriptstyle W}\right] \; ,
\nonumber \\
&+& \left( \begin{array}{cc}
 M_{\!\scriptscriptstyle Z}^2\,  \cos\!^2 \beta \;
[1 - \sin\!^2 \theta_{\!\scriptscriptstyle W}] 
& \quad 0_{\scriptscriptstyle 1 \times 3} \quad \\
0_{\scriptscriptstyle 3 \times 1} & 0_{\scriptscriptstyle 3 \times 3}  
\end{array} \right) \; ,
\nonumber \\
\widetilde{\cal M}_{\!\scriptscriptstyle RR}^2 &=&
\tilde{m}_{\!\scriptscriptstyle {E}}^2 +
m_{\!\scriptscriptstyle E} m_{\!\scriptscriptstyle E}^\dag
+ M_{\!\scriptscriptstyle Z}^2\, \cos\!2 \beta 
\left[  - \sin\!^2 \theta_{\!\scriptscriptstyle W}\right] \; ,
\nonumber \\
\widetilde{\cal M}_{\!\scriptscriptstyle H\!u}^2 &=&
\tilde{m}_{\!\scriptscriptstyle H_{\!\scriptscriptstyle u}}^2
+ \mu_{\!\scriptscriptstyle \za}^* \mu_{\scriptscriptstyle \za}
+ M_{\!\scriptscriptstyle Z}^2\, \cos\!2 \beta 
\left[ \,\frac{1}{2} - \sin\!^2\theta_{\!\scriptscriptstyle W}\right]
\nonumber \\
&+&  M_{\!\scriptscriptstyle Z}^2\,  \sin\!^2 \beta \;
[1 - \sin\!^2 \theta_{\!\scriptscriptstyle W}]  \; ;
\eeqa
and
\beqa \label{ERL}
(\widetilde{\cal M}_{\!\scriptscriptstyle RL}^{2})^{\scriptscriptstyle T} 
&=& \left(\begin{array}{c} 
0  \\   A^{\!{\scriptscriptstyle E}} 
\end{array}\right)
 \frac{v_{\scriptscriptstyle 0}}{\sqrt{2}}
- \left(\begin{array}{c} 
0  \\   m_{\!{\scriptscriptstyle E}} 
\end{array}\right)
 \, \mu_{\scriptscriptstyle 0}^* \, \tan\!\beta 
- (\, \mu_i^*\lambda_{i{\scriptscriptstyle \zb}k}\, ) \; 
\frac{v_{\scriptscriptstyle u}}{\sqrt{2}} \; ,
\\
\label{ERH}
\widetilde{\cal M}_{\!\scriptscriptstyle RH}^2
&=&  -\,(\, \mu_i^*\lambda_{i{\scriptscriptstyle 0}k}\, ) \; 
\frac{v_{\scriptscriptstyle 0}}{\sqrt{2}} \; ,
\\
\label{ELH}
\widetilde{\cal M}_{\!\scriptscriptstyle LH}^2
&=& (B_{\za}^*)  
+ \left( \begin{array}{c} 
{1 \over 2} \,
M_{\!\scriptscriptstyle Z}^2\,  \sin\!2 \beta \,
[1 - \sin\!^2 \theta_{\!\scriptscriptstyle W}]  \\
0_{\scriptscriptstyle 3 \times 1} 
\end{array} \right)
\;  .
\eeqa
Here, $m_{\!\scriptscriptstyle L}
 ={\rm diag} \{ 0, m_{\!\scriptscriptstyle E}\}\equiv
{\rm diag} \{0, m_{\scriptscriptstyle 1},
m_{\scriptscriptstyle 2},m_{\scriptscriptstyle 3}\}$, where the three
$m_i$'s are masses from leptonic Yukawa terms as discussed above
in relation to Eq.(\ref{mc}); and, again, 
$(\, \mu_i^*\lambda_{i{\scriptscriptstyle \zb}k}\, )$ denotes a matrix
($4\times 3$) with elements given by $(\;)_{{\scriptscriptstyle \zb}k}$.
Recall that for the small $\mu_i$ domain
we focused on here in this paper, we have 
$m_{\!\scriptscriptstyle E}\simeq{\rm diag} \{m_e, m_\mu, m_\tau\}$.
In fact, the $k$-th element in the 3-column-vector
$\widetilde{\cal M}_{\!\scriptscriptstyle RH}^2$ in Eq.(\ref{ERH}) 
can be written as simply as $\mu_k^* m_k$ (no sum). Similarly, the
$k$-th element in the first row of the $4\times 3$ matrix
$(\widetilde{\cal M}_{\!\scriptscriptstyle RL}^{2})^{\scriptscriptstyle T}$
in Eq.(\ref{ERL}) can be written as
$\mu_{k}^* m_k\, \tan\!\beta$ (no sum). The former is a 
$\tilde{\ell}_{\scriptscriptstyle R}^c h_{\scriptscriptstyle u}^{\mbox{-}}$
type, while the latter a 
$\tilde{\ell}_{\scriptscriptstyle R}^c h_{\scriptscriptstyle d}^{\mbox{-}}$ 
type ($h_{\scriptscriptstyle d}^{\mbox{-}} \equiv 
\tilde{\ell}_{\scriptscriptstyle L_0}$), mass-squared term. 
Or, to better illustrate the common flavor 
structure, one can put the full $F$-term part of Eq.(\ref{ERL}) as
$-\,(\, \mu_{\scriptscriptstyle \za}^*
\lambda_{{\scriptscriptstyle \za\zb}k}\, ) \, 
\frac{v_{\scriptscriptstyle u}}{\sqrt{2}}$.

For the sake of completeness, we also give explicitly  the neutral scalar,
(or sneutrino-Higgs) mass-squared matrix.
The neutral scalar mass terms, in terms of the
$(1+4)$ complex scalar fields,  $\phi_n$'s, can be written in two parts
--- a simple $({\cal M}_{\!\scriptscriptstyle {\phi}}^2)_{mn} \,
\phi_m^\dag \phi_n$ part, and a Majorana-like part in the form 
${1\over 2} \,  ({\cal M}_{\!\scriptscriptstyle {\phi\phi}}^2)_{mn} \,
\phi_m \phi_n + \mbox{h.c.}$. As the neutral scalars are originated
from chiral doublet superfields, the existence of the Majorana-like
part is a direct consequence of the electroweak symmetry
breaking VEV's, hence restricted to the scalars playing the Higgs
role only. They come from the quartic terms of the Higgs fields in
the scalar potential. We have explicitly
\beqa \label{Mpp}
{\cal M}_{\!\scriptscriptstyle {\phi\phi}}^2 =
{1\over 2} \, M_{\!\scriptscriptstyle Z}^2\,
\left( \begin{array}{ccc}
 \sin\!^2\! \beta  &  - \cos\!\beta \, \sin\! \beta
& \quad 0_{\scriptscriptstyle 1 \times 3} \\
 - \cos\!\beta \, \sin\! \beta & \cos\!^2\! \beta 
& \quad 0_{\scriptscriptstyle 1 \times 3} \\
0_{\scriptscriptstyle 3 \times 1} & 0_{\scriptscriptstyle 3 \times 1} 
& \quad 0_{\scriptscriptstyle 3 \times 3} 
\end{array} \right) \; ;
\eeqa
and
\beqa 
{\cal M}_{\!\scriptscriptstyle {\phi}}^2 &=&
\left( \begin{array}{cc}
\tilde{m}_{\!\scriptscriptstyle H_{\!\scriptscriptstyle u}}^2
+ \mu_{\!\scriptscriptstyle \za}^* \mu_{\scriptscriptstyle \za}
+ M_{\!\scriptscriptstyle Z}^2\, \cos\!2 \beta 
\left[-\frac{1}{2}\right]   
& - (B_\za) \\
- (B_\za^*) &
\tilde{m}_{\!\scriptscriptstyle {L}}^2 
+ (\mu_{\!\scriptscriptstyle \za}^* \mu_{\scriptscriptstyle \zb})
+ M_{\!\scriptscriptstyle Z}^2\, \cos\!2 \beta 
\left[ \frac{1}{2}\right]\end{array} \right) 
\nonumber \\
\nonumber \\
&+&  {\cal M}_{\!\scriptscriptstyle {\phi\phi}}^2 \; ,
\label{MSN}
\eeqa
Note that ${\cal M}_{\!\scriptscriptstyle {\phi\phi}}^2$ here is 
real (see the appendix), while 
${\cal M}_{\!\scriptscriptstyle {\phi}}^2$ does have complex entries.
The full $10\times 10$ (real and symmetric) mass-squared matrix for 
the real scalars is then given by
\beq
{\cal M}_{\!\scriptscriptstyle S}^2 =
\left( \begin{array}{cc}
{\cal M}_{\!\scriptscriptstyle SS}^2 &
{\cal M}_{\!\scriptscriptstyle SP}^2 \\
({\cal M}_{\!\scriptscriptstyle SP}^{2})^{\!\scriptscriptstyle T} &
{\cal M}_{\!\scriptscriptstyle PP}^2
\end{array} \right) \; ,
\eeq
where the scalar, pseudo-scalar, and mixing parts are
\beqa
{\cal M}_{\!\scriptscriptstyle SS}^2 &=&
\mbox{Re}({\cal M}_{\!\scriptscriptstyle {\phi}}^2)
+ {\cal M}_{\!\scriptscriptstyle {\phi\phi}}^2 \; ,
\nonumber \\
{\cal M}_{\!\scriptscriptstyle PP}^2 &=&
\mbox{Re}({\cal M}_{\!\scriptscriptstyle {\phi}}^2)
- {\cal M}_{\!\scriptscriptstyle {\phi\phi}}^2 \; ,
\nonumber \\
{\cal M}_{\!\scriptscriptstyle SP}^2 &=& -
2\, \mbox{Im}({\cal M}_{\!\scriptscriptstyle {\phi}}^2) \; ,
\eeqa
respectively. If $\mbox{Im}({\cal M}_{\!\scriptscriptstyle {\phi}}^2)$
vanishes, the scalars and pseudo-scalars decouple from one another and 
the unphysical Goldstone mode would be found among the latter. Finally, we
note that the $B_\za$ entries may also be considered as a kind of $LR$ 
mixings. The RPV $B_i$'s do in fact contribute to neutrino mass, 
as discussed below.  

We would like to emphasize that the above scalar mass results are complete 
--- all RPV contributions, SUSY breaking or otherwise, are included 
without theoretical bias. The simplicity of the result is a consequence of 
the SVP. Explicitly, there are no RPV $A$-term contributions due to the 
vanishing of VEV's $v_i\equiv \sqrt{2}\langle\hat{L}_i\rangle$.
However, such new contributions, as well as their roles in
the physics of neutrino masses and phenomena like
fermion EDM's and $b \to s \,\gamma$, are genuine. For instance, rotating
to a basis among the $\hat{L}_\za$ superfields under which the $\mu_i$'s
are zero would restore the $\hat{L}_i$ VEV's and show {\it e.g.} the RPV
$(\,\mu^*\lambda\,)$ term as a term involving the latter VEV's and some
A-term parameters. The Higgs-slepton
results given as in Eqs.(\ref{ME}) and (\ref{MSN}) are admittedly not very
useful for doing scalar physics. They contain a redundancy of parameters 
and hide the unphysical Goldstone state. 
However, for the purpose of analyzing the neutrino mass contributions
as done below, they serve their purpose. Hence, we will refrain from 
further laboring on the algebra here. 

Before ending the section, we want to emphasize the following. While it should 
be straight forward to write down the scalar mass matrices by the complete
theory of SUSY without R-parity, under any formulation or parametrization,
to the best of our knowledge, this has not been published before. After
completing the work, we checked the literature and found no explicitly
written down complete results for ${\cal M}_{\!\scriptscriptstyle {D}}^2$ and
${\cal M}_{\!\scriptscriptstyle {E}}^2$ as given here. Especially  the
existence of the interesting new RPV contributions, of the type given by
the $(\,\mu^*\lambda^{\!\prime}\,)$ term in 
${\cal M}_{\!\scriptscriptstyle {D}}^2$, and the $(\,\mu^*\lambda\,)$  term in 
Eq.(\ref{ERL}), and the $\mu^*\mu$ terms in Eqs.(\ref{ME}) and (\ref{MSN}), 
if noticed, have not been much appreciated. Explicit discussion of terms
of the type $(\,\mu^*\lambda\,)$ and their contribution to neutrino masses
was first given in a recent paper by K. Cheung and the present 
author\cite{as1}, in a different context. Their existence and important
phenomenological implication seems, otherwise, to have been overlooked.

We are not aware of any other explicit discussion of the $(\,\mu^*\lambda^{\!\prime}\,)$ and $(\,\mu^*\lambda\,)$ terms in the literature. What follows is a check into the literature for the scalar mass 
results. Before doing that, however, we would like to emphasize that the 
complete theory of SUSY without R-parity is not a very popular subject,
compared to various versions of more specific (assumed) forms of
R-parity violation. We also warn the readers that most of the previous authors 
were not working under the parametrization we used here. In our opinion, 
there are many other subtle complications when a different 
parametrization is used, which have not been explicitly addressed. The
interested readers are referred to a forthcoming review by the author on the 
subject\cite{as8}. The complete scalar mass results in a generic 
$\hat{L}_\za$ flavor basis would look more complicated than what we have
here too. In particular, there would be contributions involving
the trilinear RPV $A$-terms. Having said that, let us take a look at some 
works on a more or less complete version of R-parity violation and the 
scalar mass expression given therein.
Ref.\cite{cckl} gives ``sfermion mass matrix"  exactly as in the
MSSM without RPV terms at all.  Ref.\cite{kias} is a more careful and
detailed study. However, as mentioned above, the paper considers only one 
$\lambda$ and one $\lambda^{\!\prime}$ and hence does not give result in 
the complete theory anyway. An admissible $(\,\mu^*\lambda^{\!\prime}\,)$ 
term is still missing in the down-squark mass-squared matrix given.
An $8\times 8$ matrix corresponding to ${\cal M}_{\!\scriptscriptstyle {E}}^2$ 
is indeed presented. The matrix seems correct, under the starting
assumption, though the admissible $(\,\mu^*\lambda\,)$ term is not
explicitly shown. Ref.\cite{H} is the first to give the matrix in the 
$8\times 8$ form, but we find no clear 
sign of the $(\,\mu^*\lambda\,)$ term. The squark mass-squared 
matrix is not explicitly given there. In fact, the paper is based on a 
specific high energy scenario, which is hence not totally generic. 
For instance, early in the paper, the soft SUSY breaking part of the 
Lagrangian is already given in a simplified form with the high energy
assumptions put in, hence not for the generic complete theory we discussed
here. The neutral scalar mass, corresponding to 
${\cal M}_{\!\scriptscriptstyle {S}}^2$ above, is better known\cite{ns}. 
In particular, Ref.\cite{GH} gives the result essentially under the SVP,
though truncated to one lepton family.

Another important point to note is that we have pay special attention
to the fact there the RPV parameters are generally complex, while this phase 
information has not been explicitly given previously. As mentioned above,
the $\lambda^{\!\prime}$ term, for example, contributes to neutron EDM
through its imaginary part. The other example is the existence of
scalar-pseudo-scalar mixing as a result of complex RPV parameters, as given
above explicitly in ${\cal M}_{\!\scriptscriptstyle {SP}}^2$.
The significant phenomenological implication of the latter has been
well illustrated in Ref.\cite{PW}, in the case of MSSM, for instance.

Finally, we note that there is
another group that has done quite elaborated works on their RPV model
(see for example Ref.\cite{HDPRV}), which however includes no trilinear
RPV parameters and hence would not have our results for the complete theory.
Moreover, in the soft SUSY breaking part given and used in Ref.\cite{HDPRV}
actually has $\tilde{m}_{\!\scriptscriptstyle {L}_{0k}}^2$ term missing.

We have not actually done a detailed term by term checking to see if 
there are other discrepancies between ours results given in this section
and others in the papers mentioned. The above brief comparison is 
supposed to serve as an illustration of the point we want to make here
--- that perhaps insufficient care and attention have been given to
the complete results for the scalar masses in SUSY without R-parity. 
We will report more on the various phenomenological implications
of some of the RPV mass entries in some other publications.

\section{Neutrino Mass Contributions }
Let us return to RPV contribution to neutrino mass. From Eq.(\ref{77}),
one neutrino state get a tree-level mass. The seesaw suppressed contribution
[see Eqs.(\ref{mnu}) and (\ref{ss})] is given by 
\beq
(m_\nu)_{ij}^{\mbox{\tiny tree}} \sim \frac{- v^2 \cos^2\!\!\beta \;
( g_{\scriptscriptstyle 2}^2 M_{\scriptscriptstyle 1} 
+ g_{\scriptscriptstyle 1}^{2} M_{\scriptscriptstyle 2} )}
{2 \mu_{\scriptscriptstyle 0} \;
[2 \mu_{\scriptscriptstyle 0} M_{\scriptscriptstyle 1} 
M_{\scriptscriptstyle 2} - v^2 \sin\!\beta \cos\!\beta\; 
(g_{\scriptscriptstyle 2}^2 M_{\scriptscriptstyle 1} 
+ g_{\scriptscriptstyle 1}^{2}M_{\scriptscriptstyle 2} ) ]} \;\;
\mu_i \mu_j \;.
\eeq
This is illustrated diagrammatically in Fig.~1.
\FIGURE{\epsfig{file=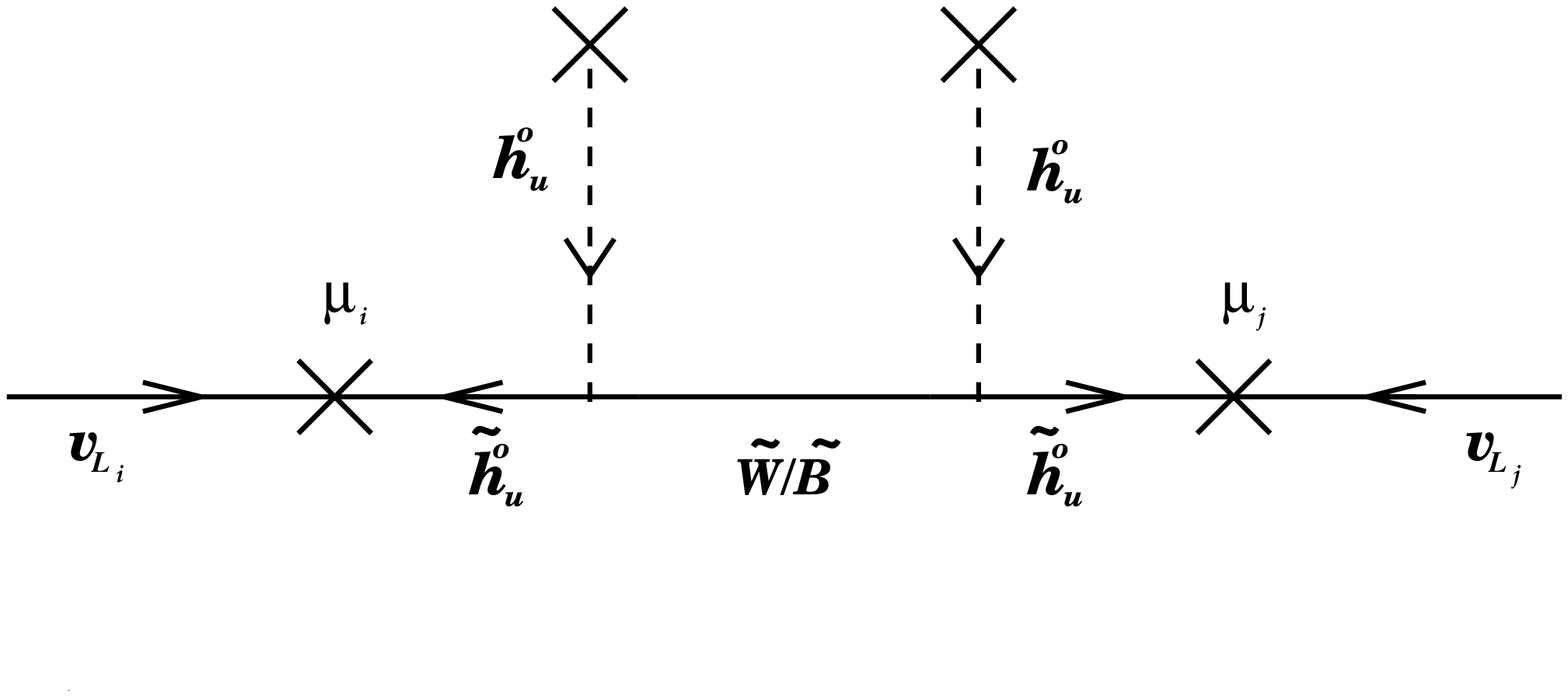, angle=270, width=5in} 
\caption{Neutrino mass from tree-level seesaw.}
}

Next we come to the direct 1-loop contributions. A typical 1-loop neutrino
mass diagram has two couplings of scalar-fermion-neutrino type. With
the two couplings being $\lambda^{\!\prime}$-type, 
we have a quark-squark loop as shown in Fig.~2. Here, a $LR$ squark mixing 
is needed. From Eqs.(\ref{MD}) and (\ref{RL}), we have the result, here
written in three parts : firstly the familiar one 
\beq 
(m_\nu)_{ij}^{\mbox{\tiny sqA}} \sim 
\frac{3}{16\pi^2}\,  \,
\frac{m_{\!\scriptscriptstyle D_h}m_{\!\scriptscriptstyle D_k}}
{M_{\scriptscriptstyle \tilde{d}}^2}\;
\lambda^{\!\prime}_{ihk} \lambda^{\!\prime}_{jkh}\;
\left[ A_d - \mu_{\scriptscriptstyle 0}^* \, \tan\!\beta 
 \right] \qquad (\; i\;\longleftrightarrow\;j\;) \; ,
\eeq
where $M_{\scriptscriptstyle \tilde{d}}$ denote an average down-squark mass,
and $A_d$ being a constant (mass) parameter representing the 
``proportional" part of the $A$-term, namely
$A^{\!{\scriptscriptstyle D}} \frac{v_{\scriptscriptstyle 0}}{\sqrt{2}}
= A_d \,m_{\!\scriptscriptstyle D} +\delta\! A^{\!{\scriptscriptstyle D}} 
\; \frac{v_{\scriptscriptstyle 0}}{\sqrt{2}}\;$,
and $m_{\!\scriptscriptstyle D_h}$ is the $h$-th diagonal element of 
the matrix $m_{\!\scriptscriptstyle D}$ ({\it i.e.} the quark mass);
next, the ``proportionality" violating part
\beq 
(m_\nu)_{ij}^{\mbox{\tiny sq$\delta\!$A}} \sim 
\frac{3}{16\pi^2}\,  \,
\frac{m_{\!\scriptscriptstyle D_h}}
{M_{\scriptscriptstyle \tilde{d}}^2}\;
\lambda^{\!\prime}_{ihl} \lambda^{\!\prime}_{jkh}\;
\left[ \delta\!A^{\!{\scriptscriptstyle D}}_{kl} 
\frac{v_{\scriptscriptstyle 0}}{\sqrt{2}} 
 \right] \qquad (\; i\;\longleftrightarrow\;j\;)\; ,
\eeq
\FIGURE[t]{\epsfig{file=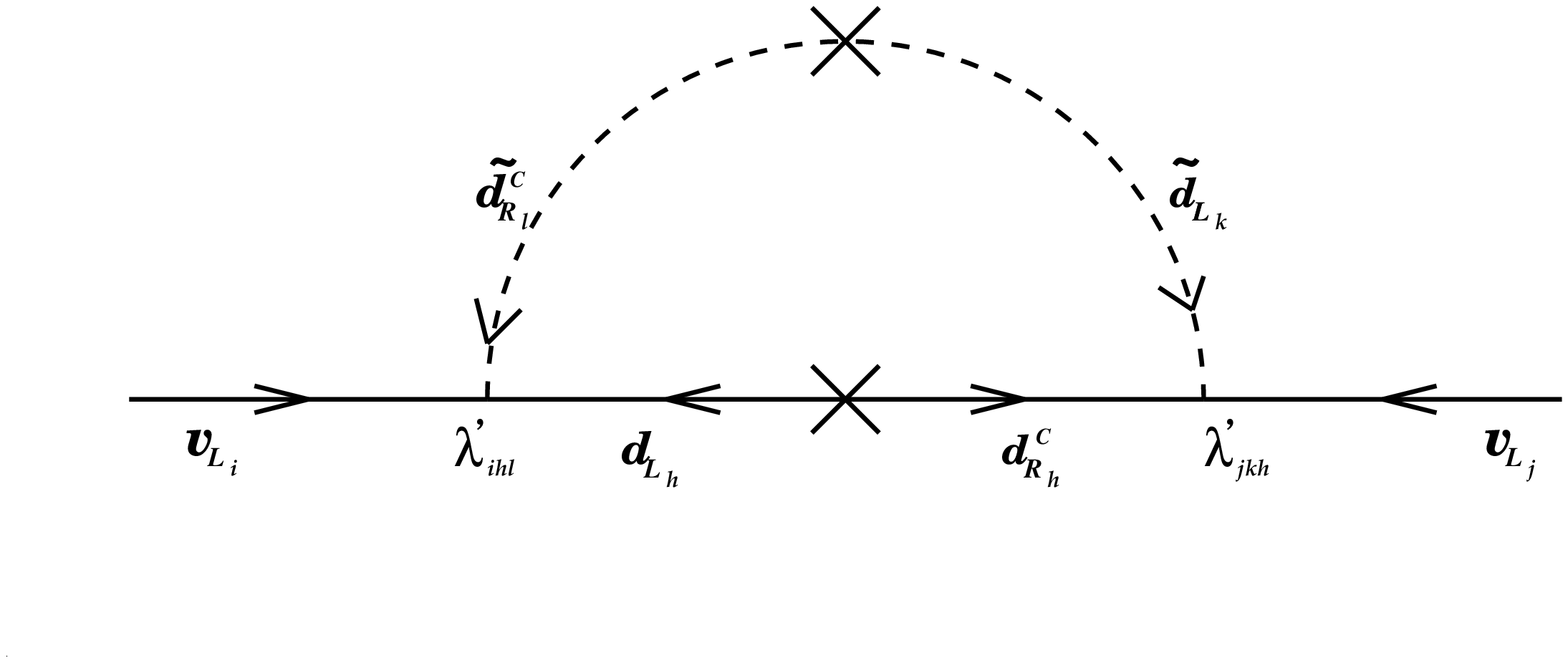, angle=270, width=5in} 
\caption{Neutrino mass from quark-squark loop.}
}
which is typically expected to be suppressed in many SUSY breaking scenarios
and neglected; and, finally, the part due to the new RPV LR mixings,
\beq 
(m_\nu)_{ij}^{\mbox{\tiny sq\Rbs}} \sim 
- \frac{3}{16\pi^2}\,  \,
\frac{m_{\!\scriptscriptstyle D_h}}
{M_{\scriptscriptstyle \tilde{d}}^2}\;
\lambda^{\!\prime}_{ihl} \lambda^{\!\prime}_{jkh}\;
\left[ \mu_g^*\lambda^{\!\prime}_{gkl} \; 
\frac{v_{\scriptscriptstyle u}}{\sqrt{2}} \right] 
\qquad (\; i\;\longleftrightarrow\;j\;)\; .
\eeq
The $(\; i\;\longleftrightarrow\;j\;)$ expression denote symmetrization
with respect to $i$ and $j$. It is interesting to note that the last 
result contains no SUSY breaking parameter in the $LR$ mixings. In 
particular, the flavor changing parts of the latter could not be suppressed
through any SUSY breaking mechanism.

Similar to the quark-squark loop, a lepton-slepton loop with two 
$\lambda$-type coupling, as shown in Fig.~3, generates neutrino mass, in
the presence of LR slepton mixings.  Using Eqs.(\ref{ME}) and (\ref{ERL}), 
again we split the result into the different parts :  the familiar
one from the ``proportional" part of the $A$-term,
\beq 
(m_\nu)_{ij}^{\mbox{\tiny slA}} \sim 
\frac{1}{16\pi^2}\,  \,
\frac{m_{h}m_{k}}
{M_{\scriptscriptstyle \tilde{\ell}}^2}\;
\lambda_{ihk} \lambda_{jkh}\;
\left[ A_e  - \mu_{\scriptscriptstyle 0}^* \, \tan\!\beta 
 \right] \qquad (\; i\;\longleftrightarrow\;j\;) \; ,
\eeq
where $M_{\scriptscriptstyle \tilde{\ell}}$ denote an average charged 
slepton mass, and $A_e$ the constant (mass) parameter with
$A^{\!{\scriptscriptstyle E}} \frac{v_{\scriptscriptstyle 0}}{\sqrt{2}}
= A_e \,m_{\!\scriptscriptstyle E} +\delta\! A^{\!{\scriptscriptstyle E}} 
\; \frac{v_{\scriptscriptstyle 0}}{\sqrt{2}}\;$, (recall that $m_{h}$'s
are diagonal element of $m_{\!\scriptscriptstyle E}$ and essentially the
mass of the charged lepton);
the ``proportionality" violating part
\beq 
(m_\nu)_{ij}^{\mbox{\tiny sl$\delta\!$A}} \sim 
\frac{1}{16\pi^2}\,  \,
\frac{m_{h}}
{M_{\scriptscriptstyle \tilde{\ell}}^2}\;
\lambda_{ihl} \lambda_{jkh}\;
\left[ \delta\!A^{\!{\scriptscriptstyle E}}_{kl} 
\frac{v_{\scriptscriptstyle 0}}{\sqrt{2}} 
 \right] \qquad (\; i\;\longleftrightarrow\;j\;)\; ;
\eeq
and the part due to the new RPV LR mixings,
\beq \label{slR}
(m_\nu)_{ij}^{\mbox{\tiny sl\Rbs}} \sim 
- \frac{1}{16\pi^2}\,  \,
\frac{m_{h}}
{M_{\scriptscriptstyle \tilde{\ell}}^2}\;
\lambda_{ihl} \lambda_{jkh}\;
\left[ \,\mu_g^*\lambda_{gkl} \; 
\frac{v_{\scriptscriptstyle u}}{\sqrt{2}} \right] 
\qquad (\; i\;\longleftrightarrow\;j\;)\; .
\eeq
\FIGURE[t]{\epsfig{file=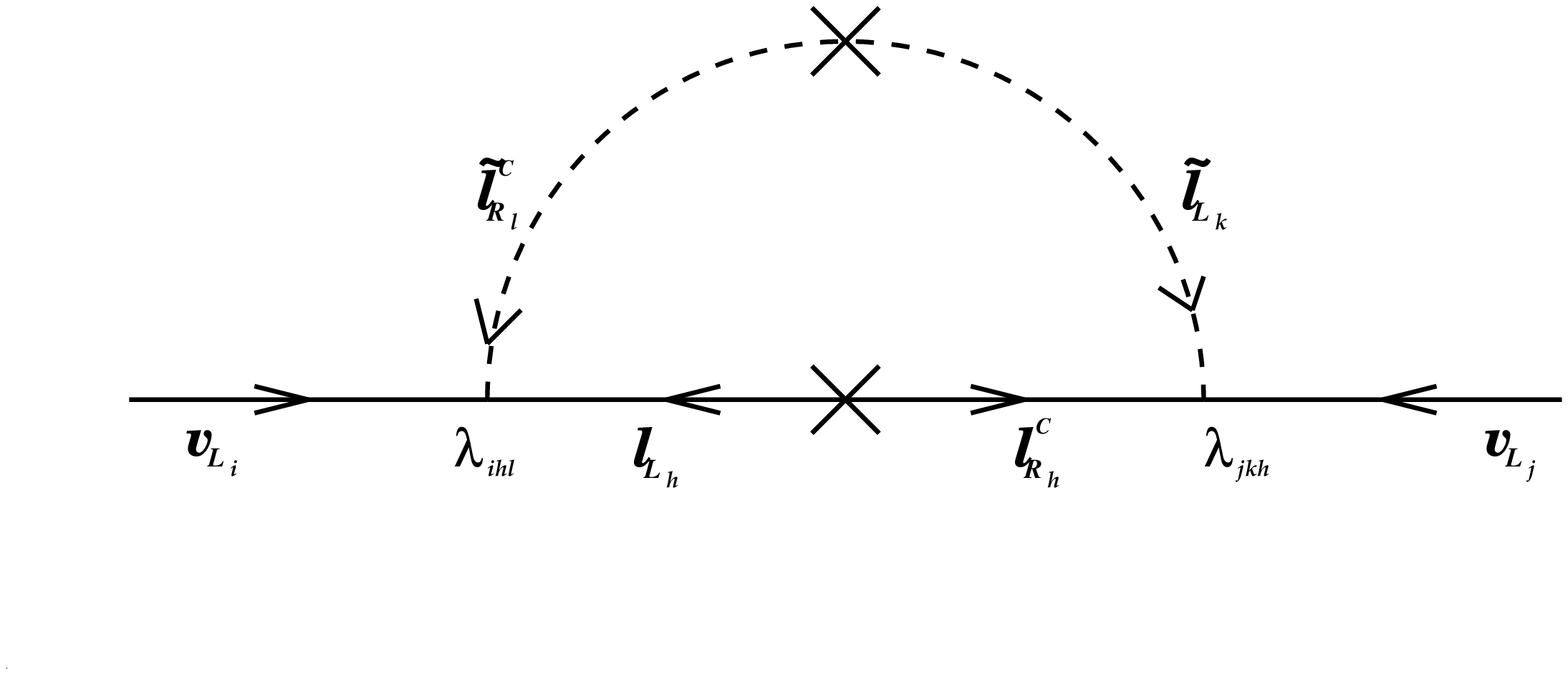, angle=270, width=5in} 
\caption{Neutrino mass from lepton-slepton loop.}
}

However, the above is not yet the full result for the type of contributions.
We have emphasized throughout the paper the systematic treatment of
making no {\it a priori} distinction between the $\hat{L}_i$'s and
$\hat{H}_d$. The latter is denoted as $\hat{L}_0$ and treated as a 4-th
leptonic flavor. In Eq.(\ref{ERL}), the last term admitted a $\zb=0$
part the neutrino mass contribution of which has not been included in 
the above analysis of the lepton-slepton loop parallel to the 
quark-squark loop. The corresponding result is simply given by setting 
$k$ to $0$ in Eq.(\ref{slR}), which may then be simplified to
\beq 
(m_\nu)_{ij}^{\mbox{\tiny slZ}} \sim 
 -\frac{1}{16\pi^2}\, \frac{\sqrt{2}}{v_{\scriptscriptstyle 0}} \,
\frac{m_{j}^2}
{M_{\scriptscriptstyle \tilde{\ell}}^2}\;
\lambda_{ijl} \;
\left[\, \mu_{l}^* m_l\, \tan\!\beta \, \right] 
\qquad (\; i\;\longleftrightarrow\;j\;)\; .
\eeq
\FIGURE[t]{\epsfig{file=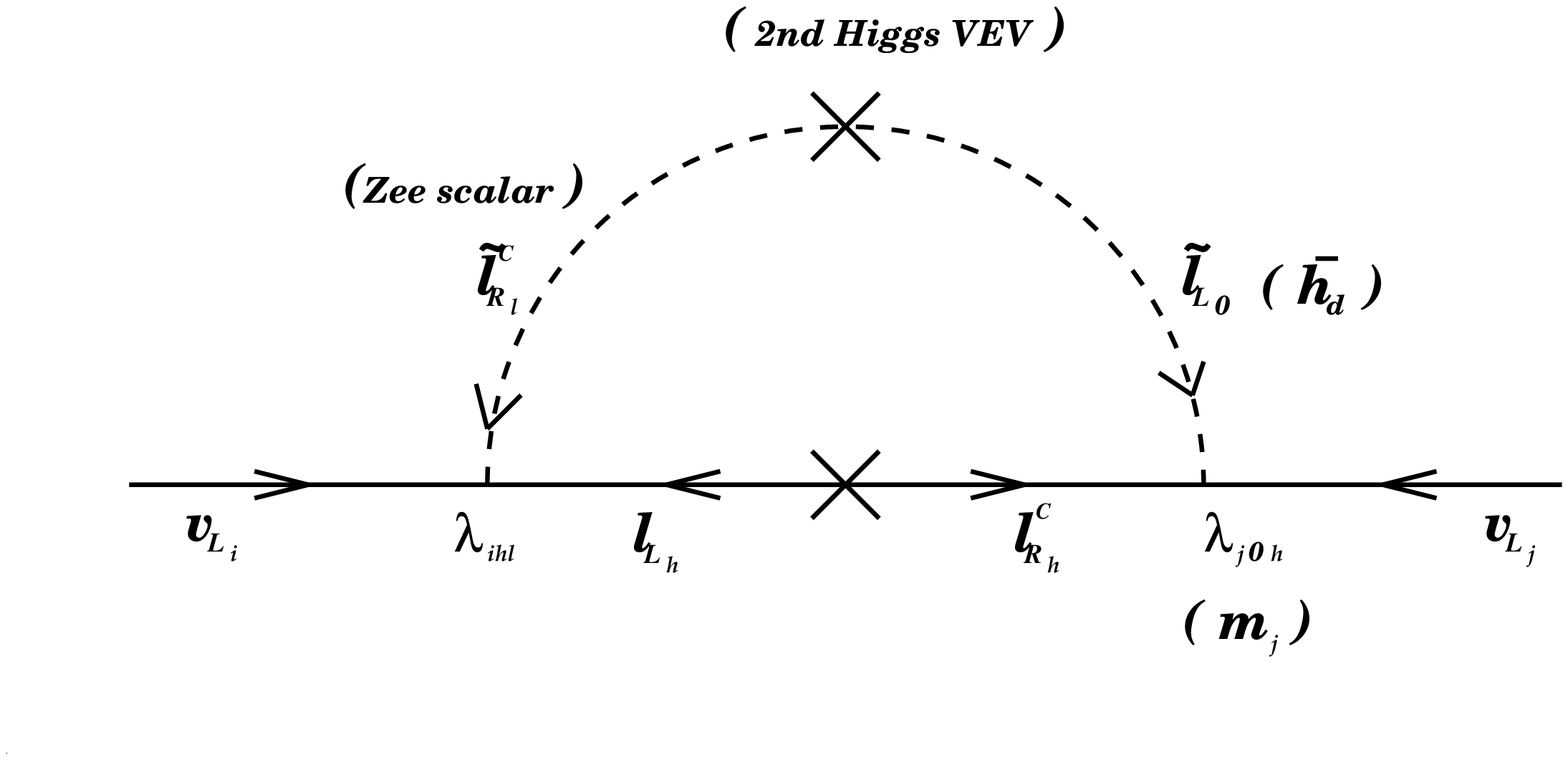, angle=270, width=5in} 
\caption{SUSY Zee diagram for neutrino mass. The Zee analog 
interpretation noted in the brackets.}
}
The contribution corresponds to the SUSY analog of the 
Zee neutrino mass diagram\cite{zee}, as discussed in Ref.\cite{as1}. 
We illustrate the contribution and its Zee model analog in Fig.~4.
A careful examination of Fig.~3 shows that one cannot get any more new
neutrino mass diagram by replacing some other $\lambda_{ijk}$ flavor 
indices with a $0$. Hence, we have completed the listing of
the two-$\lambda$-loop contributions.

The above has exhausted all the possibilities from using from the trilinear
superpotential couplings as the only source for the loop vertices.
The only other couplings involving a neutrino are the gauge couplings 
and the bilinear $\mu_i$'s. The effect of the latter has been considered
in the tree-level seesaw. Putting two gauge couplings together, we do
have a 1-loop neutrino mass diagram, with scalars and gauginos running
in the loop. The charged loop does not work, while a neutral loop could 
do (see Fig.~5) when there is a Majorana-like sneutrino mass term.
The latter contribution was first pointed out in Ref.\cite{GH}. 
In fact, Majorana-like sneutrino mass is where the required two units of
lepton number violation come in. The former may be interpreted as a 
result of splitting in mass of the sneutrino and anti-sneutrino  due to 
R-parity violation. Following our general approach here, we illustrate 
this in Fig.~6. It is clear from the figure that it involves the 
SUSY breaking and RPV parameters $B_i$'s, as shown is Eq.(\ref{MSN})
above, and is seesaw suppressed ({\it cf.} Fig~1), unless the $B_i$'s
happen to be at the SUSY scale despite small $\mu_i$'s. Note that the
$B_i$'s and the $\mu_i$'s are not totally independent parameters, as
illustrated in the appendix. It is very unlikely that 
${B_i \over B_0}$ would be much larger than 
$\mu_i \over \mu_{\scriptscriptstyle 0}$.
 
\FIGURE{\epsfig{file=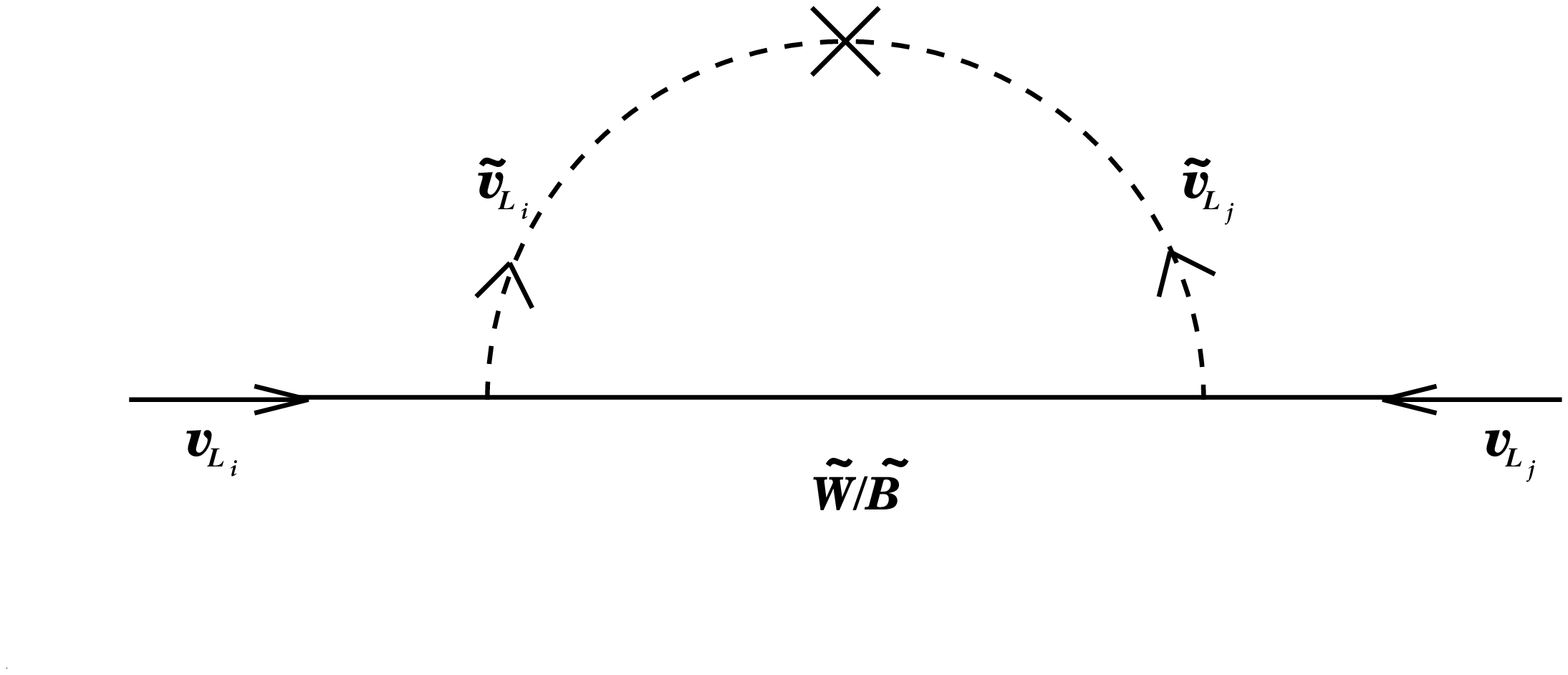, angle=270, width=5in} 
\caption{Gaugino-sneutrino loop requiring a Majorana-like
sneutrino mass insertion.}
}

\FIGURE{\epsfig{file=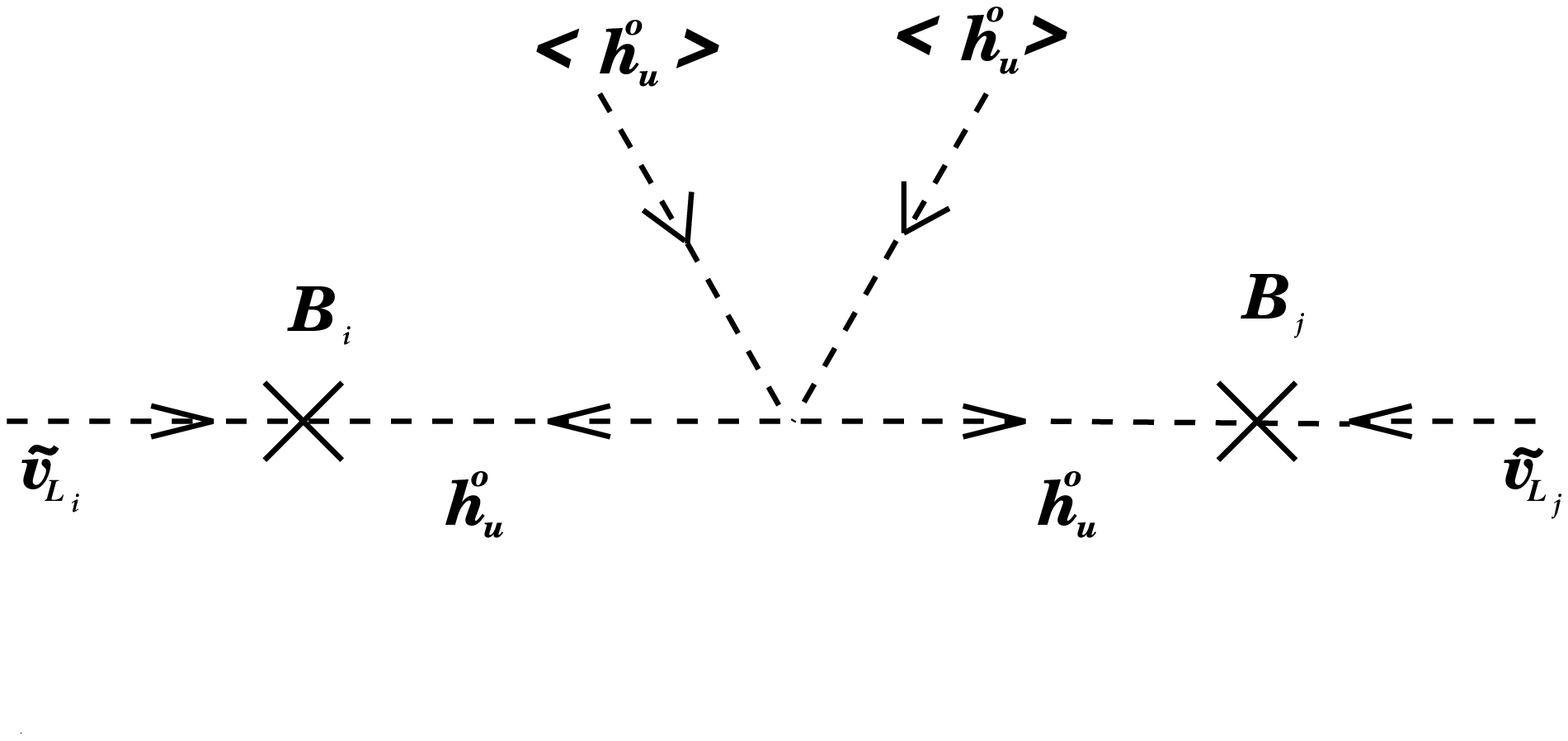, angle=270, width=5in} 
\caption{Seesaw diagram for Majorana-like sneutrino masses.}
}

The gauge loop contribution discussed serves as an illustrative example
of what we call pseudo-direct 1-loop. The EW state diagram as given in 
Fig.~5 reads zero, as the required mass insertion on the sneutrino line
does not exist [{\it cf.} Eq.(\ref{Mpp})]. If we admit extra mass insertions
and extend the sneutrino line as shown in Fig.~6, we obtain the nonzero
result. However, one show bear in mind that there is no definite hierarchy
between this gauge loop contribution and the other direct 1-loop ones
discussed above, as they arise from different RPV parameters, the magnitude
of which we have no exact information. 

Finally, it is not difficult to see that there is no contribution from 
1-loop diagrams with one gauge coupling and one Yukawa or $\lambda$-coupling
vertices, up to the direct 1-loop level. In fact, before we put in 
non-minimal number of mass insertions in the internal lines, there is only 
one such EW state diagram, as given in Fig.~7. The diagram requires a
$\tilde{W}^{\!\!\mbox{ -}}$-$\ell_{\scriptscriptstyle R_k}^{c}$
mass insertion, which is zero under the SVP. When we admit extra mass 
insertions along the internal fermion line and go to the  pseudo-direct
1-loop level, there are apparent nonzero contribution. Obviously we need
at least one lepton number violating mass insertion. Fig.~8 illustrates
the minimal extra mass insertions along the fermion line that could
complete a diagram. 

\FIGURE[t]{\epsfig{file=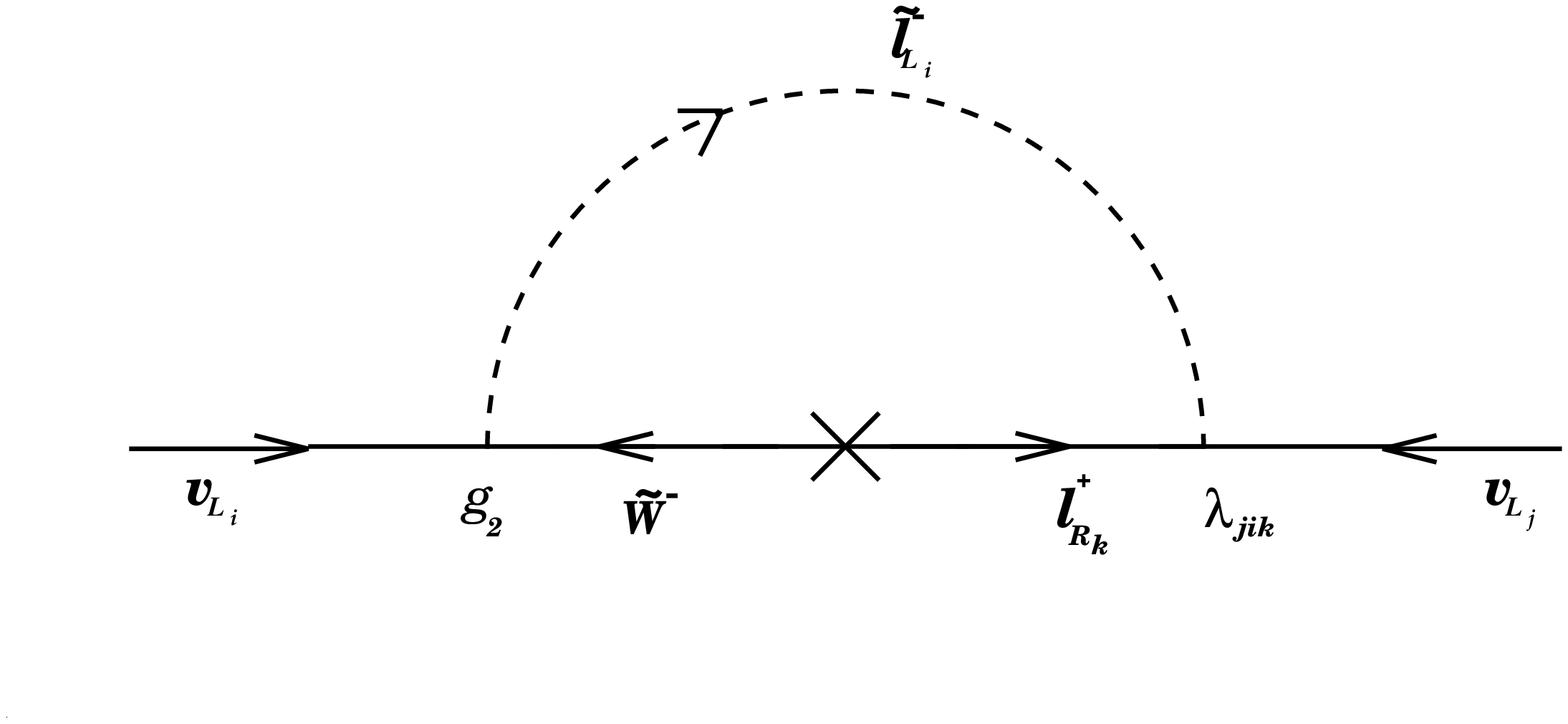, angle=270, width=5in} 
\caption{A direct charged loop contribution requiring however a
fermion mass insertion that is vanishing.}
}

\FIGURE[t]{\epsfig{file=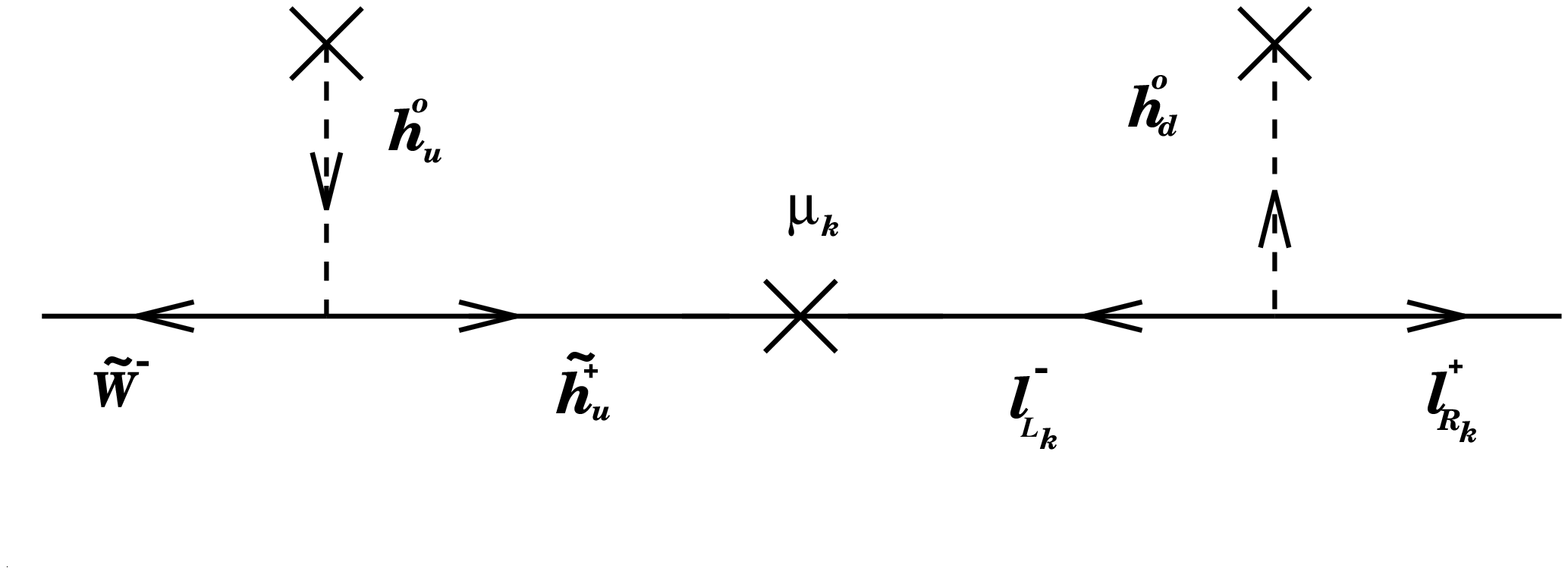, angle=270, width=5in} 
\caption{A possible set of mass insertions apparently producing the
required internal fermion line in the previous figure.}
}

However, let us look more closely into the implications
of putting in extra mass insertions along the internal fermion line. The
scalar-fermion loop neutrino mass diagram result always has a mass factor
of the internal fermion in it; explicit examples are
$m_{\!\scriptscriptstyle D_h}$ in Fig.~2 and $m_{h}$ in Fig.~3.
To obtain the exact result, one should use the mass eigenstates, and of 
course sum over the latter (see, for example, formulae in 
Ref.\cite{kias}). However, neglecting the fermion mass dependence in 
the propagator integral part, the mass and mixing matrix element 
dependence of the full sum is of course nothing other than the 
tree-level mass entry of single insertion case, {\it i.e.} the vanishing
$\tilde{W}^{\!\!\mbox{ -}}$-$\ell_{\scriptscriptstyle R_k}^{c}$ 
term for the case at hand. If one consider only the contribution from
one of the mass eigenstates, the result would not be zero. But we know that
it is going to be canceled by that from the other mass eigenstates.
This is like the GIM mechanism, violated here only to the extent of the 
non-universal mass effect from the propagator integral part.

In terms of EW state diagrams, the exact mass eigenstate result 
would correspond to summing over all possible diagrams with any 
(up to infinite) number of admissible mass insertions. The contribution
from putting Fig.~8 into the internal fermion line of Fig.~7 is 
of course just one among the nontrivial result. Hence, taking this as
an independent contribution is more or less equivalent to taking one
term from the summation over mass eigenstates. The result of the latter is
expected to be canceled in the overall sum.

Furthermore, even if that GIM-like cancellation is rendered ineffective by 
the propagator integral part, the dependence of the 
contribution on a single $\lambda_{jik}$ implies that upon symmetrization
with respect to $i$ and $j$, there would be another cancellation as
from $\lambda_{ijk}  = -\lambda_{jik}$ to the extent that 
$\tilde{\ell}_{\scriptscriptstyle L_j}^{\!\!\mbox{ -}}$ and
$\tilde{\ell}_{\scriptscriptstyle L_i}^{\!\!\mbox{ -}}$ has the same
mass. So, even the naive pseudo-direct 1-loop contribution 
has suppression from the expected degeneracy of slepton masses, and 
is only proportional to the violation of the latter. 

This, together with the gauge loop discussed above, has exhausted our
discussion of the  pseudo-direct 1-loop contributions. One could 
certainly get other EW state diagrams by putting in extra mass insertion,
into the scalar line of Figs.~5 or 7. While direct 1-loop results
from an EW state diagram roughly represent the corresponding exact mass 
eigenstate results, pseudo-direct 1-loop diagrams may be just pieces
of an otherwise small or vanishing overall sum with a GIM-like cancellation 
mechanism at work. We do have to go to exact numerical calculations
to know the extent to which the latter is violated and extract the correct
result. Perhaps it should also be said that we have not considered diagrams
with mass insertion(s) on the external lines because such diagrams really
correpond the indirect 1-loop contributions. A diagram with one mass
insertion in one of the $\nu$ external line, for example, should correspond
to something in the $\xi$ block of Eq.(\ref{77}). We want to emphasize that
while we classified contributions into direct 1-loop, pseudo-direct 1-loop
and indirect 1-loop, there is no definite hierarchy among them, when 
one is comparing contributions involving different RPV parameters. A clear
example is given by the fact that we do not know if the tree-level
contribution which involves the $\mu_i$'s is really larger than, for example
a $\lambda^{\!\prime}$ (quark-squark) loop contribution, though one may
naively expect so.

\section{Concluding Remarks }
From the above systematic analysis, it is clear that we have discussed 
and given explicit formulae for all neutrino mass contributions up to
the level of direct 1-loop contribution, for the complete theory of
SUSY without R-parity.  Psuedo-direct 1-loop contributions are 
also discussed. We have also given a description of the full 
squark and slepton masses. The latter is useful for analyzing other
aspects of phenomenology, particularly those related to LR mixings such
as fermion electric dipole moment and flavor changing neutral current
processes. The successful simple description here illustrates well 
the effectiveness of the formulation (SVP) adopted.

\vspace*{.2in}\noindent{\it Note Added :}
After we posted the first version of this paper, a paper from
Davidson and Losada\cite{DL} on the subject appeared. The paper does list
results under our formulation (SVP) here and goes beyond direct or 
pseudo-direct 1-loop level. In particular, a nice discussion of the gauge loop
result is included. However, the basic approach is very different from
that of this paper, and the contributions from the RPV $LR$ scalar mixing 
are not included.  Comparing our results with theirs, there seems to be some
apparent disagreements, which we hope to address in detail in
a future publication.  

\acknowledgments  The author is in debt to S.K. Kang 
and K. Cheung for discussions and for suggestions on improving the present 
manuscript, and to A. Akeroyd for proof-reading the last version and helps
to improve the language. K. Cheung is especially to be thanked for pointing
out some erroneous numerical factors in a couple of equations now fixed.
S. Davidson and M. Losada are to be thanked for communications
concerning their recent paper\cite{DL}.
The Korea Institute for Advanced Study is to be thanked for their
hospitality during the early phase of our work. The Yukawa Institute 
for Theoretical Phyiscs, Kyoto University, and KEK, where the last revision of 
the manuscript is done, is to be thanked likewise.

\bigskip

\appendix
\section{Note on the scalar potential}
In terms of the five, plausibly electroweak symmetry breaking, neutral 
scalars fields $\phi_n$, the generic (tree-level) scalar potential, as 
constrained by SUSY, can be written as :
\beqa
V_s &=& Y_{n} \left|\phi_n\right|^4 
+ X_{mn}  \left|\phi_m\right|^2 \left|\phi_n\right|^2 
+  \hat{m}^2_{n} \left|\phi_n\right|^2 
\nonumber \\
 && - ( \hat{m}^2_{\scriptscriptstyle m\!n} e^{i\theta\!_{m\!n}}
 \phi_m^{\dag} \phi_n + \mbox{h.c.}) \qquad\qquad\qquad (m < n) \; .
\eeqa
Here, we count the $\phi_n$'s from $-1$ to $3$ and identify a $\phi_\za$
(recall $\za=0$ to $3$) as 
$\tilde{l}_{\!\scriptscriptstyle \za}^{\scriptscriptstyle 0}$ 
and $\phi_{\scriptscriptstyle \mbox{-}\!1}$ as 
${h}_{\!\scriptscriptstyle u}^{\!\scriptscriptstyle 0}$.
Parameters in the above expression for $V_s$ (all real) are then given by
\beqa
 \hat{m}^2_{\za} &=& \tilde{m}^2_{\!{\scriptscriptstyle L\!_{\za\za}}}
 + \left|\mu_\za\right|^2 \; , \nonumber \\
 \hat{m}^2_{\scriptscriptstyle \mbox{-}\!1} &=& 
 \tilde{m}_{\!\scriptscriptstyle H\!_u}^2  + \mu_\za^{*} \mu_\za \; ,
\nonumber \\
\hat{m}^2_{\!\za\!\zb}\, e^{i\theta\!_{\za\!\zb}} &=& 
- \tilde{m}^2_{\!{\scriptscriptstyle L}\!_{ \za\!\zb}}
-  \mu_\za^{*} \mu_\zb \quad\quad  \mbox{(no sum)}\; ,	\nonumber \\
\hat{m}^2_{\!{\scriptscriptstyle \mbox{-}\!1\!\za}}\, 
e^{i\theta\!_{\scriptscriptstyle \mbox{-}\!1\!\za}} &=& B_\za 
\qquad\qquad\qquad\quad  \mbox{(no sum)}\; , \nonumber \\
 Y_{n} &=& 	\frac{1}{8}(g^2_{\scriptscriptstyle 1} 
 + g_{\scriptscriptstyle 2}^2)\; ,	\nonumber \\
 X_{{\scriptscriptstyle \mbox{-}\!1\!\za}}  &=& - 
\frac{1}{4}(g_{\scriptscriptstyle 1}^2 
 + g_{\scriptscriptstyle 2}^2) = -
 X_{{\scriptscriptstyle \za\!\zb}} \;.
\eeqa
Under the SVP, we write the VEV's as follows :
\beqa 
v_{\scriptscriptstyle \mbox{-}\!1}\,  (\equiv \sqrt{2}\,
\lla \phi_{\scriptscriptstyle \mbox{-}\!1} \rra)
& =& v_{\scriptscriptstyle u} \; , \nonumber \\
v_{\scriptscriptstyle 0} \, (\equiv \sqrt{2}\,
\lla \phi_{\scriptscriptstyle 0} \rra)
& =& v_{\scriptscriptstyle d}\, e^{i\theta\!_v} \; , \nonumber \\
v_{\scriptscriptstyle i} \, (\equiv \sqrt{2}\,
\lla \phi_i \rra)& =& 0 \; ,
\eeqa 
where we have put in a complex phase in the VEV $v_{\scriptscriptstyle 0}$,
for generality. 

The equations from the vanishing derivatives of $V_s$ along
$\phi_{\scriptscriptstyle \mbox{-}\!1}$ and $\phi_{\scriptscriptstyle 0}$
give
\beqa
{ \left[ \frac{1}{8}(g_{\scriptscriptstyle 1}^2  + g_{\scriptscriptstyle 2}^2) 
(v_{\scriptscriptstyle u}^2 -v_{\scriptscriptstyle d}^2) + 
\hat{m}^2_{\scriptscriptstyle \mbox{-}\!1} \right]} 
\, v_{\scriptscriptstyle u}
&=&
B_0 \, v_{\scriptscriptstyle d} \, e^{i\theta\!_v} \; ,
\nonumber \\
{ \left[ \frac{1}{8}(g_{\scriptscriptstyle 1}^2  + g_{\scriptscriptstyle 2}^2) 
(v_{\scriptscriptstyle d}^2 -v_{\scriptscriptstyle u}^2) + 
\hat{m}^2_{\scriptscriptstyle 0} \right] }
\, v_{\scriptscriptstyle d}
&=&
B_0 \, v_{\scriptscriptstyle u} \, e^{i\theta\!_v} \; .
\eeqa
Hence, $B_0 \, e^{i\theta\!_v}$ is real. In fact, the part of $V_s$ that
is relevant to obtaining the tadpole equations is no different from
that of MSSM apart from the fact that 
$\tilde{m}_{\!\scriptscriptstyle H\!_u}^2$ and 
$\tilde{m}_{\!\scriptscriptstyle H\!_d}^2$ of the latter are replaced by
$\hat{m}^2_{\scriptscriptstyle \mbox{-}\!1}$ and
$\hat{m}^2_{\scriptscriptstyle 0}$ respectively. As in MSSM, the $B_0$
parameter can be taken as real. The conclusion here
is therefore that $\theta\!_v$ vanishes, or all VEV's are real, despite
the existence of complex parameters in the scalar potential.
Results from the other tadpole equations, in a $\phi_i$ direction, are quite
simple. They can be written as complex equations of the form 
\begin{equation}
 \hat{m}^2\!\!_{{\scriptscriptstyle \mbox{-}\!1}i}\; 
e^{i\theta\!_{{\scriptscriptstyle \mbox{-}\!1}i}} \tan\!\beta
= -  e^{i\theta\!_v} \;
 \hat{m}^2\!\!_{{\scriptscriptstyle 0}i}\; 
e^{i\theta\!_{{\scriptscriptstyle 0}i}} \; ,
\end{equation}
which is equivalent to 
\begin{equation}
B_i \, \tan\!\beta 
=  \tilde{m}^2_{{\scriptscriptstyle L}_{\!{\scriptscriptstyle 0}i} }
+ \mu_{\scriptscriptstyle 0}^{*} \, \mu_i \; ,
\end{equation} 
where we have used $v_{\scriptscriptstyle u}=v\sin\!\beta$ and
$v_{\scriptscriptstyle d}=v\cos\!\beta$. Note that our $\tan\!\beta$ has 
the same physical meaning as that in the R-parity conserving case. For
instance, $\tan\!\beta$, together with the corresponding Yukawa coupling 
ratio, gives the mass ratio between the top and the bottom quark.

The three complex equations for the $B_i$'s reflect the redundance of 
parameters in a generic $\hat{L}_\za$ flavor basis. The equations also 
suggest that the $B_i$'s are expected to be suppressed, with respect to
the R-parity conserving $B_{\scriptscriptstyle 0}$, as the $\mu_i$'s are,
with respect to $\mu_{\scriptscriptstyle 0}$. They give consistence
relationships among the involved RPV parameters (under the SVP) that
should not be overlooked.

\end{document}